\documentclass[conference,a4paper]{APSIPA2021}

\usepackage{graphicx}
\usepackage[utf8]{inputenc}
\usepackage[ruled,vlined]{algorithm2e}
\usepackage{booktabs}
\usepackage{color}
\usepackage[nobreak]{cite}
\usepackage{listings}
\usepackage{url}
\usepackage{siunitx}
\usepackage[skip=3pt]{subcaption}

\usepackage{amsfonts}
\usepackage{amsmath}
\usepackage{amssymb}
\usepackage{amsthm}
\usepackage{bm}
\usepackage{diffcoeff}
\usepackage{dsfont}
\usepackage{interval}
\usepackage{latexsym}
\usepackage{mathtools}
\usepackage{mdframed}
\usepackage{verbatim}

\usepackage[pdftex]{hyperref}
\hypersetup{hidelinks,
  colorlinks=true,
  allcolors=black,
  pdfstartview=Fit,
  breaklinks=true}

\usepackage[capitalise,nameinlink]{cleveref}
\crefformat{equation}{(#2#1#3)}
\Crefformat{equation}{#2Equation~(#1)#3}
\crefformat{section}{Section~#2#1{}#3}
\crefname{algorithm}{Algorithm}{Algorithms}
\crefname{figure}{Fig.}{Figs.}
\crefname{equation}{}{}

\graphicspath{{./fig/}}

\newlength{\figwidth}
\newlength{\plotwidth}
\makeatletter
\def\bstctlcite{\@ifnextchar[{\@bstctlcite}{\@bstctlcite[@auxout]}}
\def\@bstctlcite[#1]#2{\@bsphack
  \@for\@citeb:=#2\do{%
    \edef\@citeb{\expandafter\@firstofone\@citeb}%
    \if@filesw\immediate\write\csname #1\endcsname{\string\citation{\@citeb}}\fi}%
  \@esphack}
\makeatother

\usepackage{textcomp}
\usepackage[absolute,showboxes]{textpos}

\TPMargin{5pt}

\newcommand{\copyrightstatement}{
   \begin{textblock}{0.84}(0.08,0.01)    % tweak here: {box width}(leftposition, rightposition)
      \noindent
      \footnotesize
      \textcopyright~2022~IEEE.
      Personal use of this material is permitted.
      Permission from IEEE must be obtained for all other uses,
      in any current or future media,
      including reprinting/republishing this material for advertising or promotional purposes,
      creating new collective works,
      for resale or redistribution to servers or lists,
      or reuse of any copyrighted component of this work in other works.
   \end{textblock}
}

\newcommand{\logdet}[1]{\log \lvert \det #1 \rvert}

\newcommand{\C}{\mathbb{C}}

\newcommand{\hermite}{\mathsf{H}}

\newcommand{\nohermite}{\vphantom{\hermite}}

\newcommand{\freq}{f}
\newcommand{\Freq}{F}

\newcommand{\tframe}{t}
\newcommand{\Tframe}{T}

\newcommand{\src}{k}
\newcommand{\Src}{K}

\newcommand{\mic}{m}
\newcommand{\Mic}{M}

\newcommand{\forget}{\alpha}

\newcommand{\ft}{\freq \tframe}

\newcommand{\source}{s}
\newcommand{\Source}{\bm{s}}

\newcommand{\Observe}{\bm{x}}

\newcommand{\estimate}{y}
\newcommand{\Estimate}{\bm{y}}

\newcommand{\srcvar}{r}

\newcommand{\cov}{\bm{U}}

\newcommand{\demix}{\bm{w}}
\newcommand{\Demix}{\bm{W}}

\newcommand{\mix}{\bm{a}}
\newcommand{\Mix}{\bm{A}}

\newcommand{\iss}{v}
\newcommand{\Iss}{\bm{\iss}}

\newcommand{\cost}{J}
\newcommand{\auxcost}{J^+}

\newcommand{\contrast}{G}%

\newcommand{\weight}{\varphi}

\newcommand{\eye}{\bm{e}}
\newcommand{\Eye}{\bm{E}}

\usepackage{interval}
\intervalconfig{soft open fences}

\newcommand{\markOne}{\textsf{`one'}}
\newcommand{\markAll}{\textsf{`all'}}
\newcommand{\propOne}{\textsf{ISS (one)}}
\newcommand{\propAll}{\textsf{ISS (all)}}
\newcommand{\convOne}{\textsf{IP (one)}}
\newcommand{\convAll}{\textsf{IP (all)}}

\title{Inverse-free Online Independent Vector Analysis\\with Flexible Iterative Source Steering}

\author{%
  \authorblockN{Taishi Nakashima and Nobutaka Ono}
  \authorblockA{%
    Tokyo Metropolitan University, Tokyo, Japan.\\
    E-mail: \href{mailto:nakashima-taishi@ed.tmu.ac.jp}{nakashima-taishi@ed.tmu.ac.jp}, \href{mailto:onono@tmu.ac.jp}{onono@tmu.ac.jp}
  }
}

\begin{document}
\maketitle
\thispagestyle{empty}
\bstctlcite{IEEE:BSTcontrol}

% copyright block for arXiv preprint
\copyrightstatement

\begin{abstract}
In this paper, we propose a new online independent vector analysis (IVA) algorithm for real-time blind source separation (BSS).
In many BSS algorithms, the iterative projection (IP) has been used for updating the demixing matrix, a parameter to be estimated in BSS.
However, it requires matrix inversion, which can be costly, particularly in online processing.
To improve this situation, we introduce iterative source steering (ISS) to online IVA.
ISS does not require any matrix inversions, and thus its computational complexity is less than that of IP.
Furthermore, when only part of the sources are moving, ISS enables us to update the demixing matrix flexibly and effectively so that the steering vectors of only the moving sources are updated.
Numerical experiments under a dynamic condition confirm the efficacy of the proposed method.
\end{abstract}

\section{Introduction}
Blind source separation (BSS) is a signal processing technology that estimates original sources from observed mixtures without prior information~\cite{Makino:2007:BSS,Makino:2018:ASS}.
For example, separating the target speech from background noise or other speech is desirable in hearing aid systems.

Many BSS methods have been proposed.
For example, independent vector analysis (IVA)~\cite{Kim:2006:ASLP,Hiroe:2006:ICA} is one of the extensions of ICA and models higher-order dependencies between frequency components.
Auxiliary-function-based IVA (AuxIVA) \cite{Ono:2011:WASPAA} realizes fast and stable estimation of demixing matrices based on the auxiliary function technique~\cite{Hunter:2004:SIAM}.

In recent years, many methods have been proposed for real-time BSS that extend batch algorithms to online ones.
It has been reported that online AuxIVA~\cite{Taniguchi:2014:HSCMA} is faster and more stable than online IVA with the gradient method.
Also, there are many extensions of online AuxIVA, such as low-latency BSS with truncation of the noncausal components~\cite{Sunohara:2017:ICASSP}, joint optimization with dereverberation~\cite{Ueda:2021:ICASSP}, and an extension of the source model to semisupervised nonnegative matrix factorization~\cite{Wang:2021:IS}.
These methods employ the iterative projection (IP)~\cite{Ono:2011:WASPAA,Kitamura:2016:ASLP} to update the demixing matrix.
IP requires matrix inversion, which can be costly.
However, reducing computational costs in many applications, especially hearing aid devices or embedded systems, is crucial because their computational power is limited.
The iterative source steering (ISS)~\cite{Scheibler:2020:ICASSP} algorithm was proposed for batch AuxIVA as a faster estimation method of the demixing matrix.
ISS is an inverse-free update rule and straightforwardly applicable to other BSS methods such as independent low-rank matrix analysis~\cite{Kitamura:2016:ASLP,Nakashima:2021:ICASSP}.

In this paper, we propose a new online AuxIVA, named {\it online AuxIVA-ISS}, by combining the autoregressive estimation of weighted covariance matrices and the ISS update rules.
The proposed method does not include any matrix inversion, increasing online BSS speed.
Furthermore, as discussed in~\cite{Scheibler:2020:ICASSP}, the demixing matrix update with ISS corresponds to the update of the steering vector.
ISS enables us to update the demixing matrix flexibly when only part of the sources are moving.
For example, if only one of the $\Src$ sources is moving, IP must update all the demixing vectors.
In contrast, in ISS, it is only necessary to update the demixing matrix so that the steering vector of the moving sources is updated.
We compare the separation performance before and after the movement of a source by simulation using three speech mixtures.
It is found that the proposed AuxIVA using ISS obtained almost the same results as AuxIVA using IP but with less calculation.

\section{Problem formulation}
Let $\Src$ be the number of sources and microphones,
and
$\Source _{\ft} = \begin{bmatrix} \source _{1\ft} & \dots & \source _{\Src\ft} \end{bmatrix} ^{\top} \in \C^{\Src}$
denote the source signals in the STFT domain.
The multichannel observed signals
$\Observe _{\ft}$
are modeled as the following convolutive mixture:
\begin{equation}
  \Observe _{\ft}
    = \textstyle\sum _{\src} \mix _{\src\ft} \source _{\src\ft}
    = \Mix _{\ft} \Source _{\ft}, \label{eq:model:mix}
\end{equation}
where
$\src$,
$\freq$,
and
$\tframe$
denote channels, frequency bins, and time frames, respectively.
$\Mix _{\ft} \in \C ^{\Src\times\Src}$
is called the mixing matrix,
and its column vectors, also known as \emph{steering vectors},
$\mix _{\src\ft}\; (\src=1,\,\dots,\,\Src)$
correspond to the transfer function from the $\src$th source to each microphone.
Note that mixing matrices $\Mix _{\freq}$ are assumed to be time-invariant in many BSS methods,
whereas mixing matrices $\Mix _{\ft}$ have time frame indices $\tframe$ in this paper because mixing matrices may vary under dynamic environments.
Online BSS aims to estimate the demixing matrices
$\Demix _{\ft} = \begin{bmatrix} \demix _{1\ft} \, \dots \, \demix _{\Src\ft} \end{bmatrix} ^{\hermite} \in \C ^{\Src\times\Src}$, which should ideally be an inverse system of~\eqref{eq:model:mix},
from only the current and past observed signals $\left\{\Observe _{\ft ^{\prime}} \middle|\ \tframe ^{\prime} \leq \tframe\right\}$.
Then, the sources are estimated as
\begin{equation}
  \Estimate _{\ft} = \Demix _{\ft} \Observe _{\ft}.
\end{equation}

Henceforth, $^{\top}$, $^{\hermite}$, and $\det$ denote vector/matrix transpose, Hermitian transpose, and determinant, respectively.
$\Eye$ denotes the $\Src$-dimensional identity matrix and $\eye _{\src} \in \mathbb{R}^{\Src}$ denotes the $\src$th canonical basis vector.
Unless specified,
frequency bin index $\freq$ ranges from 1 to $\Freq$, and
time frame index $\tframe$ ranges from 1 to $\Tframe$.

\section{Related work}
\subsection{Batch AuxIVA~\cite{Ono:2011:WASPAA}}
As the basis of our work, we first summarize batch AuxIVA.
The goal of batch IVA is to estimate time-invariant demixing matrices
$\Demix _{\freq}$
from all the observations $\Observe _{\ft}\; (\forall t)$
such that
$\Estimate _{\ft} = \Demix _{\freq} \Observe _{\ft}$
is the maximum likelihood estimation of the source signal
$\Source _{\ft}$
under the following assumptions:
\begin{enumerate}
  \item the sources $\source _{\src \ft}\; (\forall\src)$ are statistically independent,
  \item the sources follow the spherical super-Gaussian distribution,
    \begin{equation*}
      p(\source _{\src 1 \tframe}, \dots, \source _{\src \ft}) \sim
      \exp \left( - \contrast \left(\sqrt{\textstyle\sum _{\freq} \lvert \source _{\src \ft} \rvert ^{2}}\right) \right),
    \end{equation*}
    where $\contrast (\srcvar)$ is called the contrast function, which is strictly increasing and differentiable with
    $\contrast '(\srcvar) / 2 \srcvar$ strictly decreasing (see \cite{Ono:2011:WASPAA,Ono:2012:APSIPA} for details).
\end{enumerate}
Under these assumptions, demixing matrices $\Demix _{\freq}\; (\forall \freq)$ can be estimated by minimizing the negative log-likelihood of the observed signal.
\begin{align}
  \cost &=
    \sum _{\src} \frac{1}{\Tframe} \sum _{\tframe} \contrast \left(\sqrt{\textstyle\sum _{\freq} \lvert \demix ^{\hermite} _{\src\freq} \Observe ^{\nohermite}_{\ft} \rvert ^2}\right)
    - 2 \sum _{\freq} \logdet{\Demix _{\freq}}
  \label{eq:cost:auxiva} \\
  \intertext{In batch AuxIVA, we consider an auxiliary function of \eqref{eq:cost:auxiva} such as}
  \auxcost &= \sum _{\freq} \left(\sum _{\src} \demix _{\src \freq} ^{\hermite} \cov _{\src \freq} \demix _{\src \freq} - 2 \logdet{\Demix _{\freq}} \right),
  \label{eq:auxcost:auxiva}
\end{align}
where
\begin{align}
  \srcvar _{\src \tframe} &= \sqrt{\textstyle\sum _{\freq} \lvert \demix ^{\hermite} _{\src \freq} \Observe ^{\nohermite}_{\ft} \rvert ^2}, \\
  \cov _{\src \freq} &= \frac{1}{\Tframe} \sum _{\tframe} \weight(\srcvar _{\src\tframe}) \Observe _{\ft} \Observe _{\ft}^{\hermite}.
\end{align}
$\cov _{\src \freq}$ is the \emph{weighted covariance matrix} of the observed signals.
The weighting function $\weight(\srcvar)$ is determined by the source model.
For example, $\weight(\srcvar)  = 1/(2\srcvar)$ for the spherical Laplace distribution or $\weight (\srcvar) = \Freq/{\srcvar ^2}$ for the time-varying Gaussian distribution is available.
Since the closed-form solution of $\Demix_{\freq}$ that minimizes \eqref{eq:auxcost:auxiva} in a general case has not yet been found \cite{Araki:2018:IWAENC,Ono:2018:ASJ:S},
we minimize the demixing vector $\demix _{\src\freq}\; (\forall \src)$ alternatively instead of the whole demixing matrix.
\begin{align}
  \demix _{\src \freq} &\gets (\Demix _{\freq} \cov _{\src \freq}) ^{-1} \eye _{\src},
  \label{eq:IPupdate1}\\
  \demix _{\src \freq} &\gets \frac{{\demix _{\src \freq}}}{{\sqrt{\demix _{\src \freq} ^{\hermite} \cov _{\src \freq} \demix _{\src \freq}}}}.
  \label{eq:IPupdate2}
\end{align}
This update rule is called the iterative projection (IP)~\cite{Kitamura:2016:ASLP} and is guaranteed to converge.

\subsection{Online AuxIVA~\cite{Taniguchi:2014:HSCMA}}
Batch AuxIVA requires all the observed signals over time frames to calculate the weighted covariance matrices $\cov _{\src\freq}$.
However, in online applications, we must estimate the demixing matrix from the current and past observations.
The point of online AuxIVA is that the covariance matrices $\cov _{\src\freq}$ are updated in every time frame $\tframe$ in an autoregressive manner:
\begin{equation}
  \cov _{\src \ft} = \forget \cov _{\src\freq(\tframe-1)} + (1 - \forget) \weight(\srcvar _{\src \tframe}) \Observe _{\ft} \Observe _{\ft}^{\hermite},
  \label{eq:cov:online}
\end{equation}
where $\forget \in \interval[open right]{0}{1}$ is called the forgetting factor.
IP can be straightforwardly applied to online AuxIVA by replacing the update rule of the covariance matrices with \eqref{eq:cov:online}.
However, the resulting online AuxIVA procedure requires matrix inversions every time frame.
To avoid the matrix inversion, a method that utilizes a matrix inversion lemma has been proposed, but both the mixing and demixing matrices must be updated consistently in every time frame
(the details are omitted due to space limitations, see~\cite{Taniguchi:2014:HSCMA}).
In the following, we refer to online AuxIVA with IP as AuxIVA-IP.

\section{Proposed method: Online {AuxIVA}-ISS}
In this section, we propose online AuxIVA-ISS by introducing the updates of the demixing matrix with ISS to online AuxIVA.

\subsection{Online implementation of ISS}
ISS~\cite{Scheibler:2020:ICASSP} is one of the most recently proposed update rules for demixing matrices.
Instead of estimating the demixing vectors $\demix _{\src\freq}$ in IP,
ISS realizes an inverse-free update by estimating the different vectors
$\Iss _{\src\freq} = \begin{bmatrix} \iss _{1 \freq} & \dots & \iss _{\src\freq} \end{bmatrix}^{\top}$,
and updates the demixing matrices as
\begin{equation}
  \Demix _{\freq} \gets \Demix _{\freq} - \Iss _{\src\freq} \demix _{\src\freq} ^{\hermite}\quad (\forall \src).
  \label{eq:iss:update}
\end{equation}
The update rule of $\iss _{\src\freq}$ is given as
\begin{equation}
  \iss _{\mic\src\freq} =
  \begin{cases}
    1 - (\demix_{\src\freq}^{\hermite} \cov _{\src\freq} \demix_{\src\freq}^{\nohermite}) ^{-\frac{1}{2}} & \text{(if $\mic =\src$)}, \\[1em]
    \dfrac{\demix_{\mic\freq}^{\hermite} \cov _{\mic\freq} \demix_{\src\freq}^{\nohermite}}{\demix_{\src\freq}^{\hermite} \cov_{\mic\freq} \demix_{\src\freq}^{\nohermite}} & \text{(otherwise)}.
  \end{cases}\label{eq:iss:v}%\\
\end{equation}

Furthermore, in the batch processing case, %by using $\estimate _{\src} = \demix _{\src\freq}^{\hermite} \Observe _{\ft}$,
the following efficient update rules of $\Iss _{\src\freq}$ and the output estimated signal $\Estimate _{\ft}$ are available:
\begin{align}
  \iss _{\mic\src\freq}
    &= \left.\left(\sum _{\tframe} \frac{\estimate _{\mic\ft} \estimate _{\src\ft}^{*}}{\srcvar _{\mic\tframe}}\right)
        \middle/
       \left(\sum _{\tframe} \frac{\lvert \estimate _{\src\ft} \rvert ^{2}}{\srcvar _{\mic\tframe}}\right)\right., \label{eq:iss:inplace:v} \\
  \Estimate _{\ft} &\gets \Estimate _{\ft} - \Iss _{\src\freq} \estimate _{\src\ft} \label{eq:iss:inplace:y}.
\end{align}

\newcommand{\SRC}{\mathcal{I}_{\src}}
Although these updates of \eqref{eq:iss:inplace:v} and \eqref{eq:iss:inplace:y} are attractive,
we cannot use them in the online processing since we need $\Demix _{\freq}$ explicitly to separate mixtures at the new frame.
However, \eqref{eq:iss:v} is still available if we estimate the weighted covariance matrices $\cov _{\src\freq}$ in the online manner of \eqref{eq:cov:online}.
Therefore, the combination of \eqref{eq:cov:online} and \eqref{eq:iss:v} is the basis of AuxIVA-ISS.
\newcommand{\niter}{\mathrm{iter}}
\newcommand{\Niter}{N_{\mathrm{iter}}}
\begin{algorithm}[t]
  % TODO: add projection back
  \DontPrintSemicolon
  \SetKwInOut{Input}{Input}\SetKwInOut{Output}{Output}
  \Input{%
    Observed mixture $\Observe _{\freq \tframe}\; (\forall \freq, \tframe$) \\
    Number of iterations per time frame $\Niter$ \\
    Forgetting factor $\forget$ \\
    Initial demixing matrices $\Demix _{\freq 0} \; (\forall \freq)$ \\
    Initial covariance matrices $\cov _{\src\freq 0} \; (\forall \src, \freq)$
  }
  \Output{Separated signals $\Estimate _{\freq \tframe}\; (\forall \freq, \tframe)$ }
  % Set the reference microphone index $n$,
  \For{$\tframe = 1,\, \dots,\, \Tframe$}{
    $\Demix _{\ft} \gets \Demix _{\freq(\tframe - 1)}\; (\forall \freq)$\;%\tcp*{Initialization}
    \For{$\niter = 1, \dots, \Niter$}{
      \For{$\src = 1,\, \dots,\, \Src$}{
        $\srcvar _{\src\tframe} \gets \sqrt{\sum _{\freq} \lvert \demix _{\src\ft} ^{\hermite} \Observe _{\ft} \rvert ^2}$\;
        $\cov _{\src} \gets \forget \cov _{\src\freq(\tframe-1)} + (1 - \forget) \weight(\srcvar _{\src\tframe}) \Observe _{\ft} \Observe _{\ft}^{\hermite}$\;
      }
      \For{$\src \in \SRC$}{
        \For{$\mic = 1,\, \dots,\, \Mic$}{
          \eIf{$\mic \neq \src$}{
            $\iss _{\mic\src\freq} \gets
                \dfrac{\demix_{\mic\ft}^{\hermite} \cov _{\mic\ft} \demix_{\src\ft}^{\nohermite}}{\demix_{\src\ft}^{\hermite} \cov_{\mic\ft} \demix_{\src\ft}^{\nohermite}}$\;
          }{
            $\iss _{\mic\src\freq} \gets 1 - (\demix_{\src\ft}^{\hermite} \cov _{\src\ft} \demix_{\src\ft}^{\nohermite}) ^{-\frac{1}{2}}$
          }
        }
        $\Demix _{\ft} \gets \Demix _{\ft} - \Iss _{\src\freq} \demix _{\src\ft} ^{\hermite}$\;
      }
    }
    $\Estimate _{\ft} = \Demix _{\ft} \Observe _{\ft}\; (\forall \freq)$\;
  }
  \caption{Online AuxIVA-ISS.}%
 \label{alg:online:auxiva:iss}
\end{algorithm}

\subsection{Flexible updates for partly moving sources}
Suppose a situation where the online estimation almost converges, but some sources start to move.
When only part of the sources is moving, only the column vectors of the mixing matrix that correspond to the moving sources are time-varying.
In the IP case, we still must update all the demixing vectors in the same way when all sources are moving because updating the rows of the demixing matrix does not update the columns of the mixing matrix.
In contrast, ISS enables us to update the demixing matrix flexibly so that the steering vectors of only the moving sources are updated.
This is because the ISS update of \eqref{eq:iss:update} is equivalent to the update of the steering vector $\mix _{\src\freq}$~\cite{Scheibler:2020:ICASSP} as
\begin{equation}
  \mix _{\src\freq} + \bm{u}_{\src\freq} = \frac{1}{1 - \iss _{\src\src\freq}} \left(\mix _{\src\freq} + \sum _{\mic \neq \src} \iss _{\mic\src\freq} \mix _{\mic\freq}\right).
\end{equation}
Then, let $\SRC$ be the set of source indices used to which \eqref{eq:iss:update} is applied.
Before convergence, $\SRC$ should comprise all the source indices $\{1,\, \dots,\, \Src\}$.
After convergence and when we know which sources are moving, $\SRC$ should include only the indices of the moving sources, and we can apply \eqref{eq:iss:update} only for $k\in\SRC$.
\Cref{alg:online:auxiva:iss} summarizes the proposed online AuxIVA-ISS.

\subsection{Another flexibility of AuxIVA-ISS}
The original online AuxIVA employs IP to update demixing vectors, and an efficient way to reduce the computational cost of IP is proposed by applying the matrix inversion lemma \cite{Taniguchi:2014:HSCMA}.
However, we must notice that, to use the efficient algorithm, the update of the covariance matrices $\cov _{\src\ft}$ must be rank-1.
It means that we have to update the demixing vectors $\demix _{\src\ft}$ in every time frame.
In contrast, AuxIVA-ISS does not need this requirement.
For example, updating demixing vectors only once every few time frames is possible.
This flexibility of AuxIVA-ISS would also save the computational cost in real-time processing. We will evaluate this in future work.

\section{Experiment}
\subsection{Setup}
We performed simulation experiments with speech signals to confirm the advantageousness of the flexible update method of online AuxIVA-ISS.
To evaluate frame-wise separation performance, we compared the segmental SDR (SegSDR) defined in the following.
Let $s(n),\, y(n)\; (n = 1, \dots, N)$ respectively be the reference and estimated signals in discrete time domain,
and their $i$th segment be
\begin{align}
  \bm{S}_i &\coloneqq \begin{bmatrix} s\left(\left(i-1\right)L+1\right) & \dots & s\left(iL\right) \end{bmatrix}, \\
  \bm{Y}_i &\coloneqq \begin{bmatrix} y\left(\left(i-1\right)L+1\right) & \dots & y\left(iL\right) \end{bmatrix},
\end{align}
where $L$ denotes the length of each segment and $i = 1, \dots,\left\lfloor \tfrac{N}{L}\right\rfloor$.
We define SegSDR as SDR at every segment, that is $\text{SDR}(\bm{S}_i, \bm{Y}_i)$.
In this experiment, we used \texttt{BSSEval v4}~\cite{Stoter:2018:LVAICA} to compute the SDR and we set the segment length $L$ to \num{32000} samples which equals to \SI{2}{\second}.

We compared two approaches.
In \markAll{}, the update rules, \eqref{eq:IPupdate1} and \eqref{eq:IPupdate2} in IP or \eqref{eq:iss:update} in ISS,  were applied for all $k=1,\, \dots,\, \Src$ throughout the observation.
In \markOne{}, the update rules were applied for all $k=1,\, \dots,\, \Src$ in the first half of the observation and then applied for one specific $k$ corresponding to the moving source in the second half.
In this paper, we assume that the moving source is known, and its automatic detection will be future work.
In the following, we refer to
online AuxIVA-IP using \markAll{} as \convAll{},
online AuxIVA-IP using \markOne{} as \convOne{},
the proposed online AuxIVA-ISS using \markAll{} as \propAll{},
and
the proposed online AuxIVA-ISS using \markOne{} as \propOne{}.
For both experiments, we set the forgetting factor $\alpha$ to \num{0.99},
the initial $\cov _{\src\freq0}$ to $\Eye\times\num{0.001}$,
the initial $\Demix _{\freq0}$ to $\Eye$, and
the number of iterations per time frame $\Niter$ to \num{2}.
After separation, the scale of the estimated signal was restored by back-projection onto the first microphone~\cite{Murata:2001:Neurocomputing}.
The sampling frequency was \SI{16}{\kilo\hertz} and the STFT was performed with a Hamming window of length \num{1024} samples (\SI{64}{\milli\second}) with half-overlap.

\begin{figure}[t]
  \centering
  \includegraphics{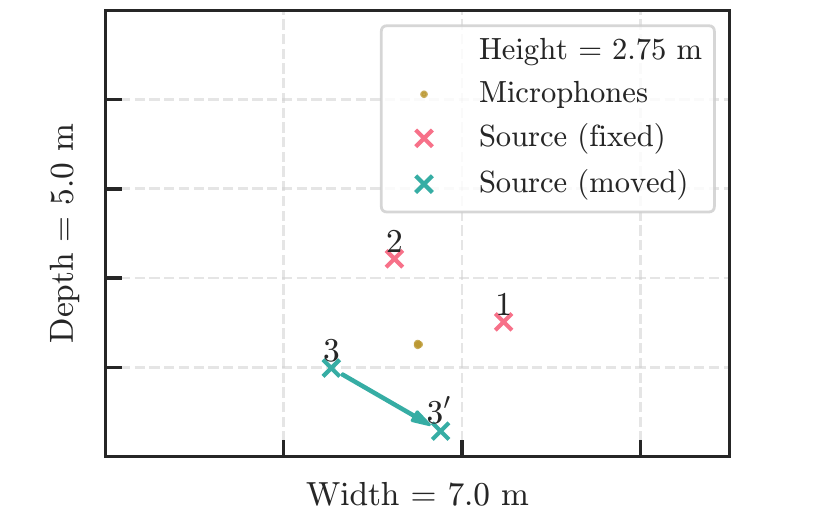}
  \caption{Simulated room layout. Source 3 instantaneously moves to $3'$ during the simulation. The microphone array and sources 1 and 2 are fixed.}%
  \label{fig:layout}
\end{figure}

We used speech signals from ASJ Japanese Newspaper Article Sentences Read Speech Corpus (JNAS)~\cite{Itou:1999:JASE} and concatenated JNAS speech signals with a length of \SI{60}{\second}.
All sources were simulated by convolving the room impulse response generated with the image source method implemented in \texttt{pyroomacoustics}~\cite{Scheibler:2018:ICASSP}.
\Cref{fig:layout} shows the layout of the simulated room.
The microphone array was circular with intervals of \SI{2}{\centi\metre}, and the reverberation time was approximately \SI{150}{\milli\second}.
As shown in \cref{fig:layout}, source \num{3} was copied to source \num{3}$'$, then source \num{3} was muted for the first \SI{30}{\second} and the source \num{3}$'$ was muted for the last \SI{30}{\second}.
This preprocess simulates a situation where source \num{3} instantaneously moves to source \num{3}$'$ at \SI{30}{\second}.

\subsection{Result}
\begin{figure}[t]
  \includegraphics{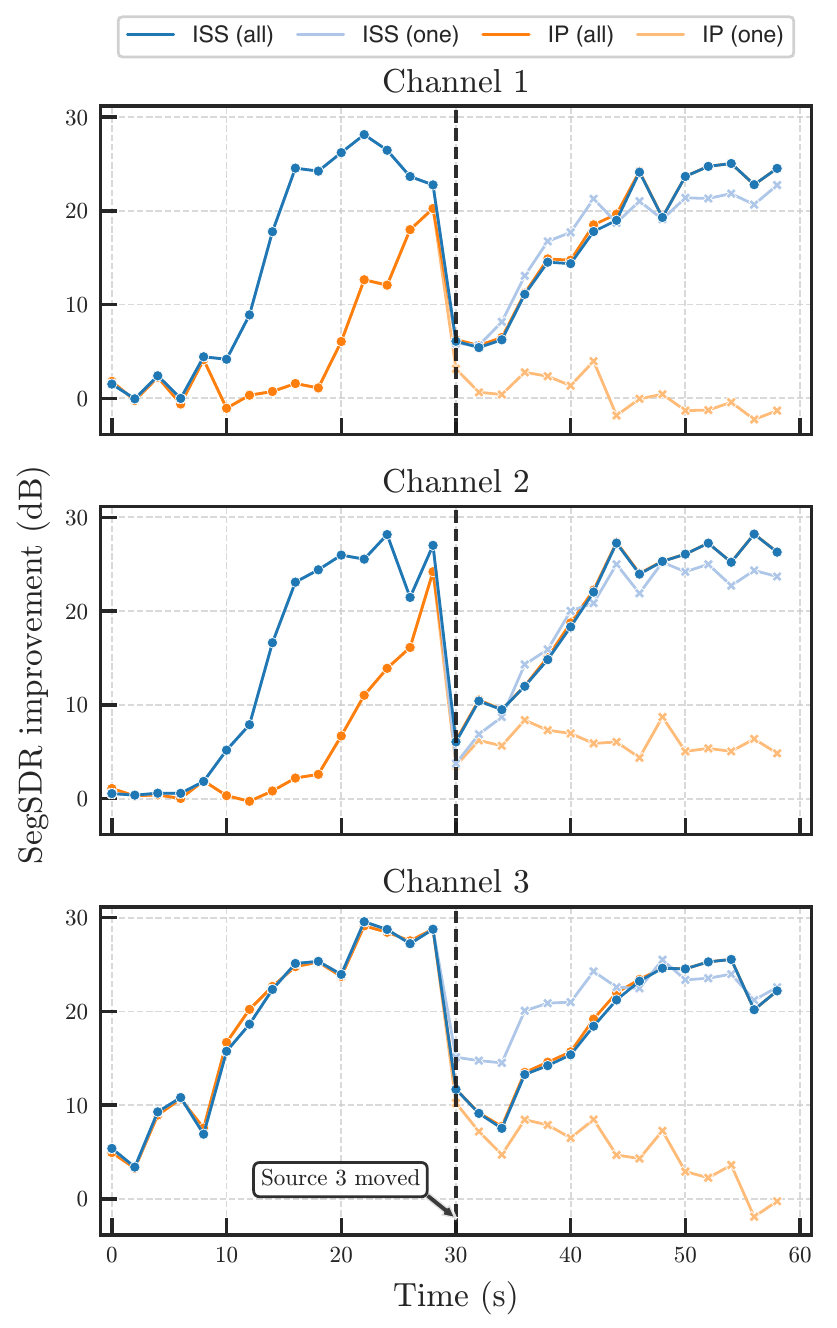}
  \caption{Improvements of segmental SDR for a mixture of three speeches. Note that the lines for \propAll{} and \convAll{} nearly overlap.}%
  \label{fig:sdr:jnas}
\end{figure}
\begin{table}[t]
  \centering
  \caption{Overall SDR improvement and runtime for a 60-second mixture of three speeches.}%
  \label{tab:sdr:runtime:jnas}
  \sisetup{%
    table-format=2.2,
    table-alignment-mode=format,
    table-auto-round=true,
    table-number-alignment=center,
    table-text-alignment=right,
  }
  \begin{tabular}{@{}lSS@{}}
    \toprule
      {Method}    & {Runtime (s)}   & {SDR improvement (dB)} \\
    \midrule
      {\propAll} & 11.001142 &  9.850487 \\
      {\propOne} & 8.660962  & 10.147417 \\
      {\convAll} & 11.544285 &  8.130473 \\
      {\convOne} & 9.004813  &  3.843095 \\
    \bottomrule
  \end{tabular}
\end{table}

\Cref{fig:sdr:jnas} shows the temporal variation of the improvement of segmental SDR.
As expected from the experimental settings, the SDR improvement of each method was significantly degraded after the source moved.
First, focusing on \propAll{} and \convAll{}, the separation performance of both methods improved over time.
Next, focusing on \propOne{} and \convOne{}, \convOne{} was not able to improve the separation performance after the source moved,
whereas \propOne{} showed the equivalent performance to \propAll{} even though only one parameter corresponding to the moved source was updated.

\Cref{tab:sdr:runtime:jnas} shows the overall SDR improvement and total runtimes.
Runtimes and SDR inprovements of \propAll{} and \convAll{} were similar.
The runtimes of \propOne{} and \convOne{} were shorter than \propAll{} and \convAll{}, thanks to the flexible update.
As for \propOne{} and \convOne{}, runtimes were comparable, but SDR improvement of \propOne{} were much better than that of \convOne{}.

\section{Conclusion}
In this paper, we proposed AuxIVA-ISS, a new online IVA.
By combining autoregressive estimation of the weighted covariance matrix and the demixing matrix update using ISS, we realized an inverse-free online algorithm.
In addition, utilizing the fact that ISS corresponds to the update of the steering vector, we proposed a flexible method of applying the ISS update only for moving sources.
Experimental results confirmed that this method could track the source movement more flexibly than IP.
We plan to develop an automatic detection of moving sources based on direction-of-arrival estimation such as~\cite{Knapp:1976:ASSP} in the future.

\section*{Acknowledgements}
This work was supported by JST CREST Grant Number \mbox{JPMJCR19A3} and JSPS KAKENHI Grant Number \mbox{JP21J22039}, Japan.

\clearpage
\newpage
\bibliographystyle{IEEEtran}
\bibliography{IEEEabrv,ref}

\end{document}